\documentclass{article}

\usepackage[numbers]{natbib}
\usepackage[preprint]{nips_2018}

\usepackage[utf8]{inputenc} 
\usepackage[T1]{fontenc}    
\usepackage{hyperref}       
\usepackage{url}            
\usepackage{booktabs}       
\usepackage{amsfonts}       
\usepackage{nicefrac}       
\usepackage{microtype}      
\usepackage{graphicx}
\usepackage[export]{adjustbox}

\MakeRobust{\overrightarrow}
\MakeRobust{\overleftarrow}

\title{Leveraging Sequence Embedding and Convolutional Neural Network for Protein Function Prediction}

\author{
    Wei-Cheng Tseng, Po-Han Chi, Jia-Hua Wu, Min Sun \\
    Department of Electrical Engineering\\
    National Tsing Hua University\\
    No. 101, Section 2, Guangfu Road, East District, Hsinchu City, Taiwan \\
    \texttt{ethanweichengtseng@gmail.com} \\
    \texttt{\{aapp1420, chacha25806\}@gapp.nthu.edu.tw}\\
    \texttt{minsun@ee.nthu.edu.tw}
}

\begin{document}

\maketitle

\begin{abstract}
    The capability of accurate prediction of protein functions and properties is essential in the biotechnology industry, e.g. drug development and artificial protein synthesis, etc. The main challenges of protein function prediction are the large label space and the lack of labeled training data. Our method leverages unsupervised sequence embedding and the success of deep convolutional neural network to overcome these challenges. In contrast, most of the existing methods delete the rare protein functions to reduce the label space. Furthermore, some existing methods require additional bio-information (e.g., the 3-dimensional structure of the proteins) which is difficult to be determined in biochemical experiments. Our proposed method significantly outperforms the other methods on the publicly available benchmark using only protein sequences as input. This allows the process of identifying protein functions to be sped up.
\end{abstract}
\section{Introduction}
    Accurate prediction of protein function is essential for biotechnology industry applications such as drug development and artificial protein synthesis. It is estimated that there are billions of different proteins in the world, but only a few of them have been discovered. Nowadays, the number of the determined protein sequences has grown rapidly. So far, there are over 120 million protein sequences in the UniProt database \citep{doi:10.1093/nar/gkw1099} with only 557 thousand of them manually annotated with protein functions (see Fig. 1). Manual annotation of the protein functions is a curation process, requiring manual integration of information from the experimental evidence. Moreover, these evidence requires experts to conduct experiments. While the results of manual annotation are more reliable and more accurate, the process is not only expensive but also time-consuming.

\begin{figure}[h]
	\includegraphics[scale=0.42]{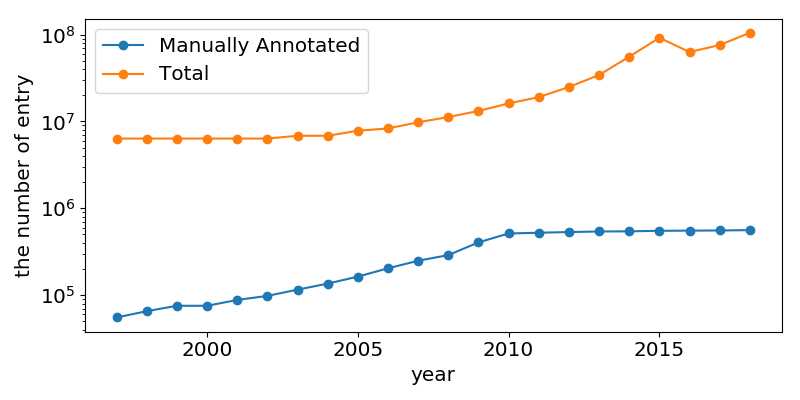}
	\centering
	\caption{The number of entries in the UniProt database. The total number of entries grows very rapidly, but the total number of manually reviewed entries doesn't.}
\end{figure}

	One alternative to the manual annotation is the computational methods, which only rely on protein sequences or protein databases. Most important of all, the computational methods enable the entire annotation process to work faster, cheaper, and automatically. As a result, the computational models offer a possible solution to greatly improve the efficiency of protein function annotation.\par

	However, several challenges with respect to prediction of protein functions need to be overcome. First, the protein functions depend on the structures of the protein. Nowadays, it still needs lots of experimental effort to measure the underlying protein structures. Second, the lengths of the protein sequences vary in a wide range. The shortest sequence contains fewer than a hundred amino acids, while the longest one contains more than thirty thousand amino acids. This phenomenon indicates that the choice of the padding length during sequence embedding has an important effect on the performance. Finally, the protein functions are annotated through Gene Ontology (GO) \citep{Ashburner2000,doi:10.1093/nar/gkw1108}. GO has defined more than 40000 terms with a hierarchy to describe functions and cellular locations, and several types of relationships between these terms. In addition, most of the proteins have multiple functions. This implies that protein function prediction is a hierarchical, multi-label classification problem. \par

    Our method takes advantage of state-of-the-art word embedding technique and parameter-wise efficient network structure. First, we regard three amino acids as a token. Therefore, an amino acid sequence is a sequence of tokens. Second, inspired by ELMo \citep{Peters:2018}, we train a language model of the amino acid sequence. The word embedding of the token is a weighted average of the hidden state of the language model. The weight of the weighted average is determined during the training process of the protein function prediction model. Third, we leverage one of the most parameter-wise efficient convolutional neural network - the inception network~\citep{szegedy2015going} - to predict the protein function given learned sequence embedding. Our method can achieve the state-of-the-art performance on UniProtKB/Swiss-Prot dataset. \par
    To sum up, our contributions are shown as following:
    
    \begin{itemize}
        \item We propose a protein function prediction model combining a sequence embedding technique and a deep convolutional neural network. Our proposed method yields state-of-the-art performance on one of largest publicly available dataset.
        \item Our proposed method has less inference time than those of the existent models. In other words, our method can accelerate the process of validation of protein functions. This implies that the efficiency of prediction of protein functions in related applications can be improved.
    \end{itemize}

\section{Relative Works}
    \subsection{Amino Acid Sequence Annotation}
        There are two types of computational method. The first type of method rely on the sequence alignment algorithm and compute the similarity between amino acid sequences. Take BLASTp \citep{BLAST} for example, this searching tool finds the regions of similarity between the amino acid  sequences. The program compares protein sequences to sequence databases and calculates the statistical significance. However, the inference speed of the alignment algorithm is too slow in most case. Pannzer \citep{doi:10.1093/bioinformatics/btu851} and Pannzer2 \citep{doi:10.1093/nar/gky350} employ a weighted k-nearest-neighbors classifier based on the sequence similarity and the enrichment statistics. However, these models simplify the problem by removing lots of GO terms which contain few amino acid sequences. For a hierarchical label space like GO, these removed terms have a generally deeper level. In conclusion, these models cannot predict many deep-level terms. \par
        The second type of methods is based on the artificial neural networks. DeepGO \citep{kulmanov2017deepgo} uses the 1D convolutional neural network to extract the local features of the amino acid sequence. However, it also uses the protein-protein interaction database as the input features, which is unavailable for the newly determined amino acid sequences.  \par

    \subsection{Sequence Embedding}
        The most well-known sequence embedding technique is word embedding. There are several outstanding word embedding representation methods in natural language processing, such as CBOW and Skip-Gram \citep{NIPS2013_5021}. Unlike the traditional representation, some of the representations are context-dependent, such as ELMo \citep{Peters:2018}. These word2vec remove the data sparsity and extract the relation between the words which improve the performance of the language model. \par
        
        Moreover, sentence embedding is a type of sequence embedding. Take skip-thought vectors \citep{DBLP:journals/corr/KirosZSZTUF15} for example, instead of predicting the words surrounding a word, the method attempts  the surroundings sentences of a given sentence. In quick-thoughts vectors \citep{logeswaran2018an}, the task of predicting the next sentence given the previous one is reformulated as a classification task: the decoder is replaced by a classifier which has to choose the next sentence among a set of candidates. It can be interpreted as a discriminative approximation to the generation problem. \par
        
        Our method leverages the state-of-the-art word embedding technique, i.e. ELMo \citep{Peters:2018}. We treat three amino acids as one token and construct a language model for the amino acid sequence. For inference and training, we feed the token into the language model and take the linear combination of the hidden state of language model as word embedding. 
    
    \subsection{Deep Neural Network}
    
        From 2013, deep neural network is proved to be a feasible solution in several fields, such as computer vision and natural language processing. VGG \citep{DBLP:journals/corr/SimonyanZ14a} perform well in image classification. GoogLeNet \citep{szegedy2015going}, ResNet \citep{DBLP:journals/corr/HeZRS15}, and DenseNet \citep{DBLP:journals/corr/HuangLW16a} improve the accuracy further. Among several famous network architecture, the inception network is parameter-wise efficient work. Given the limited training data, we use the inception network as the main architecture of our model.

\section{Background}
    Proteins are involved in virtually all cell functions, such as structure maintenance, cell signaling, and metabolism. To understand how protein functions are presented and predicted, we introduce the composition of proteins in Sec. 3.1, the relationship between a protein function and the protein sequence(s) in Sec. 3.2, and the protein annotation system in Sec. 3.3.

    \subsection{The Composition of Proteins}
        Proteins are large polymers consisting of one or multiple folded and assembled amino acid sequences. Each amino acid sequence is a linear and unbranched chain of amino acids, also known as a polypeptide. Amino acids are organic compounds containing amine (-NH$_{2}$), carboxyl (-COOH) functional groups, and residue (-R). There are 20 standard amino acids with distinct residue which varies with chemical structure and properties. While representing the amino acid sequences, the 20 standard amino acids should be written in the IUPAC single-letter codes.  \par
        Amino acid sequences are synthesized by condensation polymerization of amino acids (see Fig. 2(a)). During the polymerization process, each amino acid combines with another amino acid by a dehydration reaction between the amine group and the carboxyl group, forming a peptide bond. After polymerization, The two terminals of each chain are the amine group and the carboxyl group, called N-terminal and C-terminal, respectively. Amino acid sequences are represented by the IUPAC single-letter code from their N-terminal to C-terminal (see Fig. 2(b)). Based on the different properties among residues and their arrangement, protein functions are determined by the amino acid sequence(s). Although several types of chemical bonding ( hydrogen bond, ionic bond, and disulfide bridge)  and the folding process of the amino acid sequence also participate in protein structure formation, all of these factors are almost completely dependent on the amino acid sequence. Therefore, taking amino acid sequences as the input of the model is a feasible and reasonable choice.
        
        \begin{figure}[h]
            \includegraphics[scale=0.27]{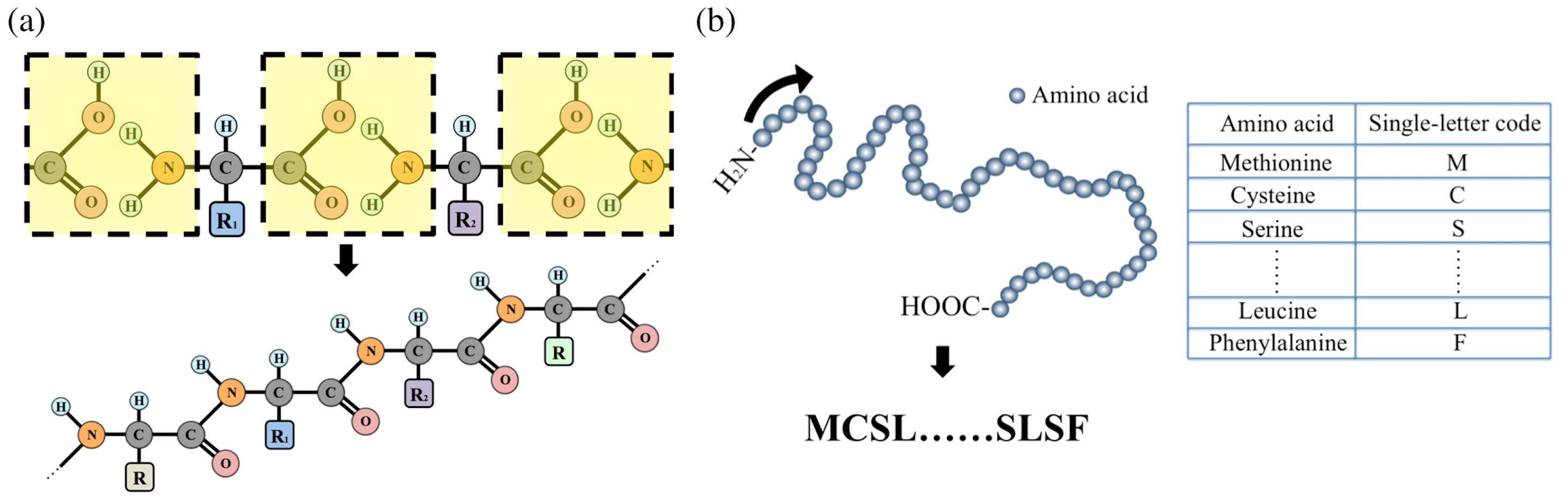}
        	\centering
        	\caption{Illustration of the amino acid polymerization and the sequence representation. (a) All the chemicals are represented by the structural formula. One amine group (-NH$_{2}$) and one carboxyl group (-COOH) react with each other (see the yellow dash-bounded boxes), forming one peptide bond (see the purple dash-bounded boxes). Amino acids within an amino acid sequence are linked by peptide bonds. (b) The two terminals are represented by the structural formula. The amino acid sequence representation begins with N-terminal and ends with  C-terminal. In addition, the 20 standard amino acids should be written in the IUPAC single-letter codes.}
        \end{figure}

    \subsection{Functions within Amino acid sequence}
        The fundamental functions of proteins are presented through the interactions between molecules. They are usually formed by one or more subsequences, such as protein domain and sequence motif. For example, the ``EF hand'' is a structural domain found in a large family of calcium-binding proteins. Each EF hand has a unique sequence pattern: two alpha-helices linked by a loop region that binds calcium ions. This kind of region can be easily recognized by analyzing patterns within sequences. However, many of the protein functions are presented through a more complicated way that is hard to be recognized by simply analyzing the sequence pattern. For this reason, the deep neural network can solve this problem by extracting deep features. 
    \subsection{Gene Ontology}
        GO defines terms for functions of the gene product, and the relationship between each term. Each term has a unique ID represented in ``GO:'', following seven digits in tens (e.g., GO:0000001). GO is widely used to annotate protein sequences in protein databases, such as UniProt. In this case, each GO term is equivalent to a label.\par
        
        Each GO term is related to one or more terms by is\_a relation, which is a subsumption relation. The hierarchical structure of GO is based on this relation. For example (see Fig. 3(a)), ``nucleus'' is\_a ``intracellular organelle'', this implies that ``nucleus'' is a subterm of ``intracellular organelle''. From another perspective, we can describe GO in terms of a graph. Therefore, each node is connected to one or more nodes by hierarchical edges, which directed from a child node to a parent node, forming a directed acyclic graph (DAG). This implies that protein function prediction is a hierarchical multi-label classification problem. Although there are several more relations defined in GO, such as part\_of and positive\_regulates, none of them can form a complete graph independently like is\_a relation. \par
        
        The entire GO consists of three domains: Cellular Component (CC), Molecular Function (MF) and Biological Process (BP). All the three domains are a complete hierarchical DAG. Moreover, the three domains are is\_a disjoint, meaning there is no is\_a relation between different domains. CC is defined as the component of a cell or an extracellular part where a gene product locates, such as the nucleus. MF is defined as the functionalities of a gene product at the molecular level, such as calcium ion binding mentioned in Sec. 3.2. This is usually formed by subsequences with specific amino acid arrangement pattern. BP is defined as a recognized series of collaborated molecular functions or chemical reactions, such as photosynthesis. All the three domains are essential for describing protein function completely. You can see an instance of the GO annotation in Fig. 3(b).  \par
        \begin{figure}[h]
        	\includegraphics[scale=0.34]{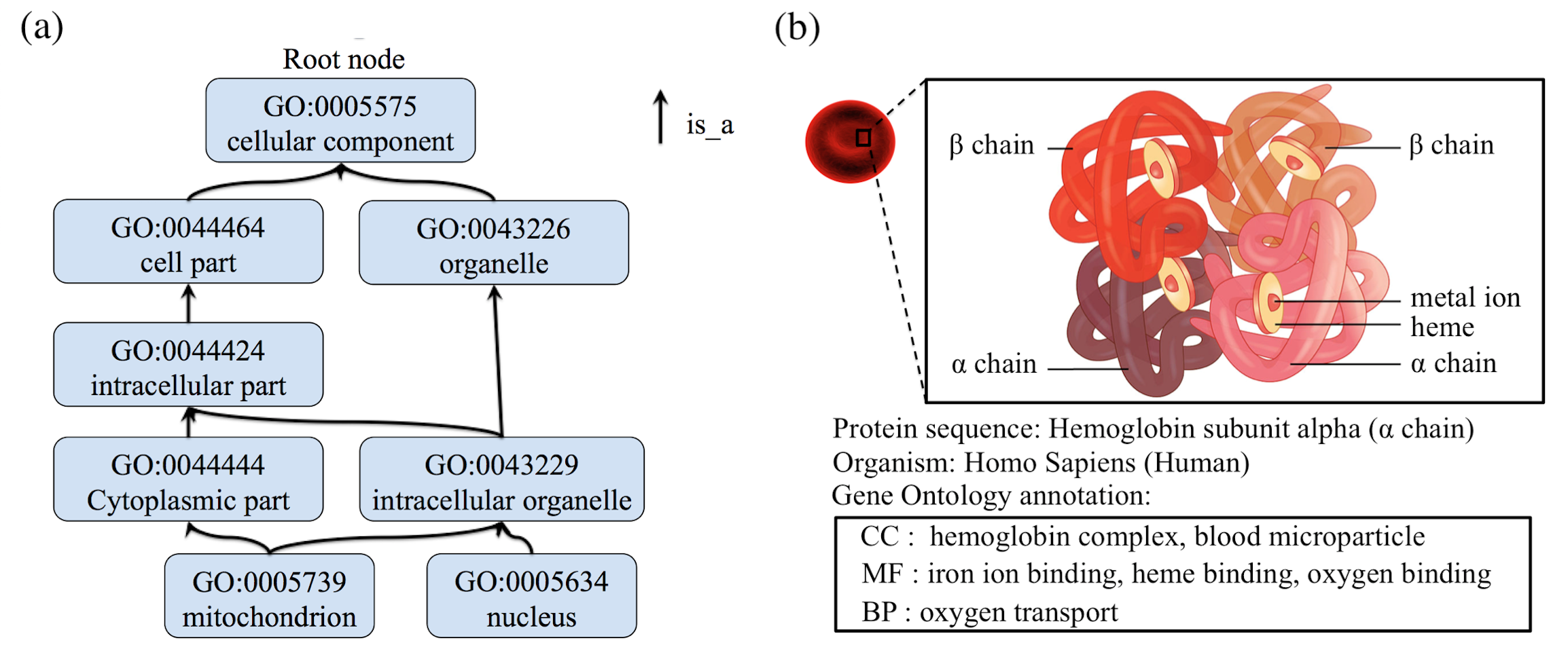}
        	\centering
        	\caption{Instance for the Gene Ontology. (a) The figure shows a part of Gene Ontology graph. We can find that GO is a hierarchical DAG with terms connected by is\_a relation, which implies a child term is\_a subterm of its parent term. (b) The above figure takes the $\alpha$ chain of human hemoglobin protein as an instance. With the GO annotation, we can learn the functions of a protein sequence by rigorously defined concepts.}
        \end{figure}
        In the 3/9/2018 version used in our experiment, there are 4171 terms in CC, 11154 terms in MF, and 29614 terms in BP, totally 44939 terms. 

\section{Our Method}
    We propose a prediction model with combination of word embedding and inception network. We first define the notation in Sec. 4.1. Then, we describe word embedding for AA sequence in Sec. 4.2, deep neural network in Sec. 4.3, loss function in Sec. 4.4. The overall architecture of our model is shown in the supplementary material.

    \subsection{Problem Definition}
        An amino acid sequence can be represented as $\mbox{S}$ = [$a_{1}$, $a_{2}$, $a_{3}$, ..., $a_{n}$], where $n$ is the length of the sequence and $a_{j}$ is an amino acid. We regard a sequence as an input of the model. Then, we can get an output vector $\mbox{Y}$ with length $q$ where $q$ is the total number of GO labels. $y_{i}$ is the probability of $i$-th GO label where $y_{i}$ $\in$ [0, 1] and $y_{i}$ $\in$ $\mbox{Y}$. Therefore, our model can be summarized as
        \begin{equation}
        \mbox{Y} = \mathbf{M}(\mbox{S}, {\phi})
        \end{equation}
        where $\mathbf{M}$ is the model we propose below and ${\phi}$ is the parameters in the model. 

    \subsection{Word Embedding}
        We regard three amino acids as one token. In other words, for an amino acid sequence $\mbox{S}$ = [$a_{1}$, $a_{2}$, $a_{3}$, ..., $a_{n}$], we map $[a_{1}$, $a_{2}$, $a_{3}]$ to $t_{1}$ and $[a_{4}$, $a_{5}$, $a_{6}]$ to $t_{2}$. In general, we map $\mbox{S}$ = [$a_{1}$, $a_{2}$, $a_{3}$, ..., $a_{n}$] into $\mbox{T}$ = [$t_{1}$, $t_{2}$, $t_{3}$, .., $t_{\lceil n/3 \rceil}$]. Since there are 20 common amino acids, we have 8000 basic protein tokens and a $<$unk$>$ which represents the other case. Moreover, $<$bos$>$ and $<$eos$>$ represent the beginning of a sequence and the end of a sequence respectively. Therefore, we have 8003 word vectors in total. In this paper, we use ELMo as our word embedding method. 

        There are two reasons to use unsupervised word embedding. First, compared with the number of the manually annotated amino acid sequences, the number of the non-annotated amino acid sequences are significantly large. Therefore, we can fully utilize the non-annotated amino acid sequences to train word embedding. Second, since the lengths of the amino acid sequences vary in a wide range, viewing three amino acids as one token can reduce the variance of the lengths of the amino acid sequences.
        
        \begin{figure}[h]
        	\includegraphics[scale=0.14]{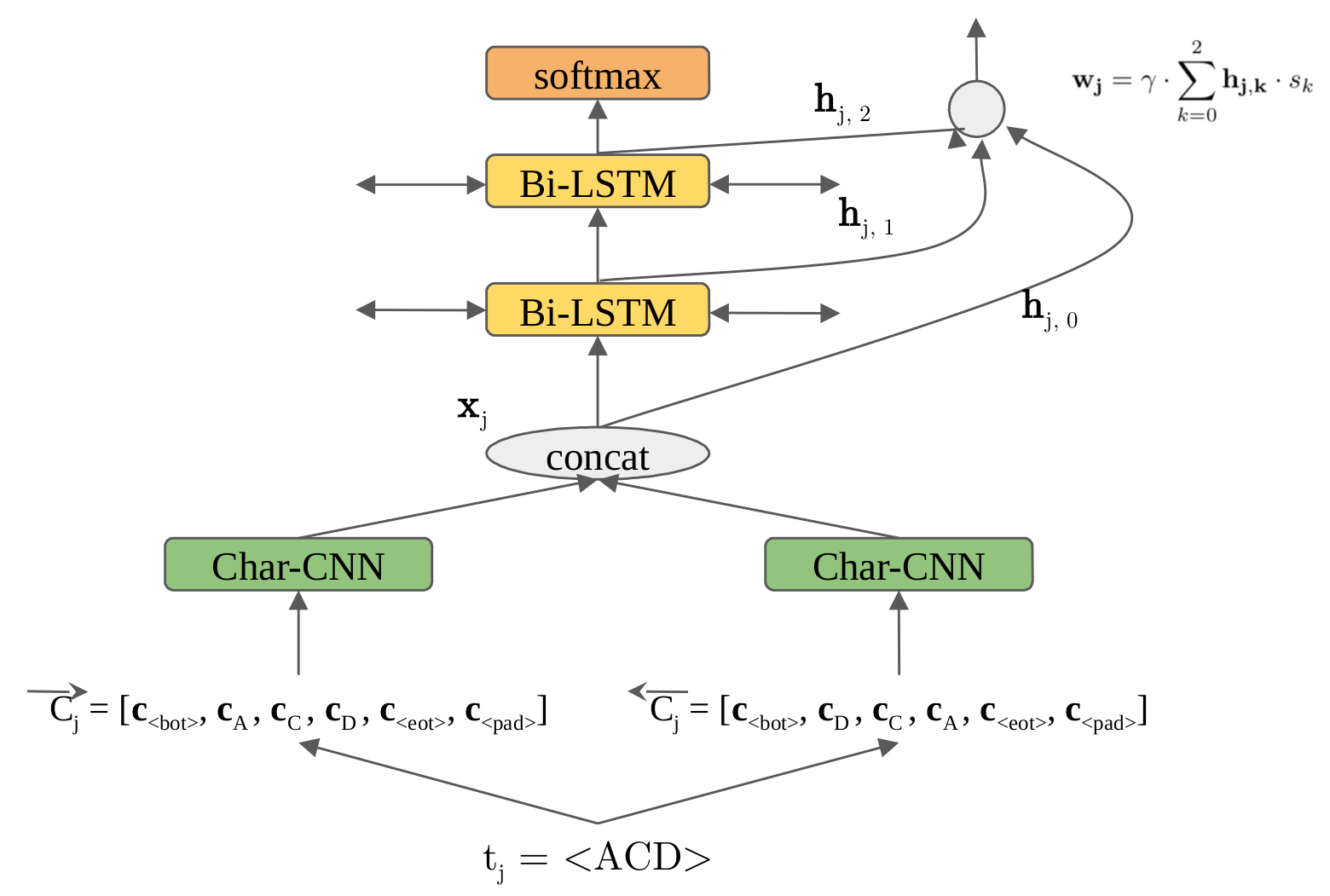}
        	\centering
        	\caption{
            	ELMo word embedding on amino sequence. For each token $t_{j}$, we first map $t_{j}$ to character matrix $\protect{\overrightarrow{\mbox{C}}}_{j}$ and $\protect{\overleftarrow{\mbox{C}}}_{j}$. Then, we feed the character matrix to char-CNN and get a context-independent token $\mathbf{x_{j}}$. $\mathbf{x_{j}}$ is passed to the Bi-LSTM model. The LSTM computes the context-dependent representation ${\mathbf{h}}_{j, k}$. Finally, we compute the weighted average of both context-independent token $\mathbf{x_{j}}$ and context-dependent representation ${\mathbf{h}}_{j, k}$ as the word vector ${\mathbf{w}}_{j}$.
        	}
        \end{figure}

        In ELMo, we first pretrain a Bi-LSTM language model for amino acid sequences (see Fig. 4). Given a sequence of $m$ tokens, $[t_{1}, t_{2}, ..., t_{m}]$, a forward language model computes the probability of the sequence by modeling the probability of token $t_{j}$ given the history $\big[t_{1}, ..., t_{j - 1}\big]$:

        \begin{equation}
            p(t_{1}, t_{2}, ..., t_{m}) = \prod_{j=1}^{m} p(t_{j}|t_{1}, t_{2}, ..., t_{j-1})
        \end{equation}

        Our language model is similar to \citep{DBLP:journals/corr/JozefowiczVSSW16,DBLP:journals/corr/KimJSR15}. We combine a CNN over characters and bi-LSTM into a language model. In our case, we regard a amino acid as a character. For a input token $t_{j} = [a_{3j}, a_{3j+1}, a_{3j+2}]$, we can get $\mbox{C}_{j} = [\mathbf{c}_{<bot>}, \mathbf{c}_{a_{3j}}, \mathbf{c}_{a_{3j+1}}, \mathbf{c}_{a_{3j+2}}, \mathbf{c}_{<eot>}, ..., \mathbf{c}_{<pad>}]$, where $\mathbf{c}_{a_{j}}$ is the character vector of $a_{j}$. We compute 1D convolution on $\overrightarrow{\mbox{C}}_{j}$ and concatenate with $\overleftarrow{\mbox{C}}_{j}$. Then, we can get context-independent vector $\mathbf{x}_{j}$. We pass $\mathbf{x}_{j}$ through $l$ layers of forward LSTM. At each position $j$, each LSTM layer outputs a context-dependent representation $\overrightarrow{\mathbf{h}}_{j,k}$ where $k$ = 1, . . . , $l$. The LSTM output, $\overrightarrow{\mathbf{h}}_{i,l}$, is used to predict the next token ${t_{j+1}}$ with a softmax layer. \par
        A backward language model is similar to a forward language model, except it runs over the sequence in reverse, and it predicts the previous token given the future context:
        
        \begin{equation}
            p(t_{1}, t_{2}, ..., t_{m}) = \prod_{j=1}^{m} p(t_{j}|t_{j+1}, t_{j+2}, ..., t_{m})
        \end{equation}
        
        Finally, after we pretrain the language model, we concatenate the forward and backward hidden state in language model, that is $\mathbf{h_{j,k}} = [\mathbf{\overrightarrow{\mathbf{h}}_{j,k}}, \mathbf{\overleftarrow{\mathbf{h}}_{j, k}}]$. Then, we compute the weighted average of both context-independent token $\mathbf{x_{j}}$ and context-dependent representation ${\mathbf{h}}_{j, k}$ as the word vector $\mathbf{w}_{j}$. Therefore, our word embedding will be 
        
        \begin{equation}
            \mathbf{w_{j}} = \gamma  \cdot \sum_{k=0}^{l} \mathbf{h_{j, k}}\cdot s_{k}
        \end{equation}
        
        where $s_{k}$ is the softmax-normalized weights, and $\mathbf{x}_{j} = \mathbf{h_{j, 0}}$. $\gamma$ allows our following protein sequence model to scale the ELMo word embedding. For detail implementation, please refer to \citep{Peters:2018}

    \subsection{Deep Neural Network}
        \subsubsection{Transition Layer}
            The transition layer is denoted by \textbf{H}. Except for \textbf{H} after the first convolution layer, all \textbf{H} are the combination of batch normalization (BN) \citep{Ioffe2015BatchNA} and elu activation \citep{DBLP:journals/corr/ClevertUH15}.

            \begin{figure}[h]
            	\includegraphics[scale=0.28]{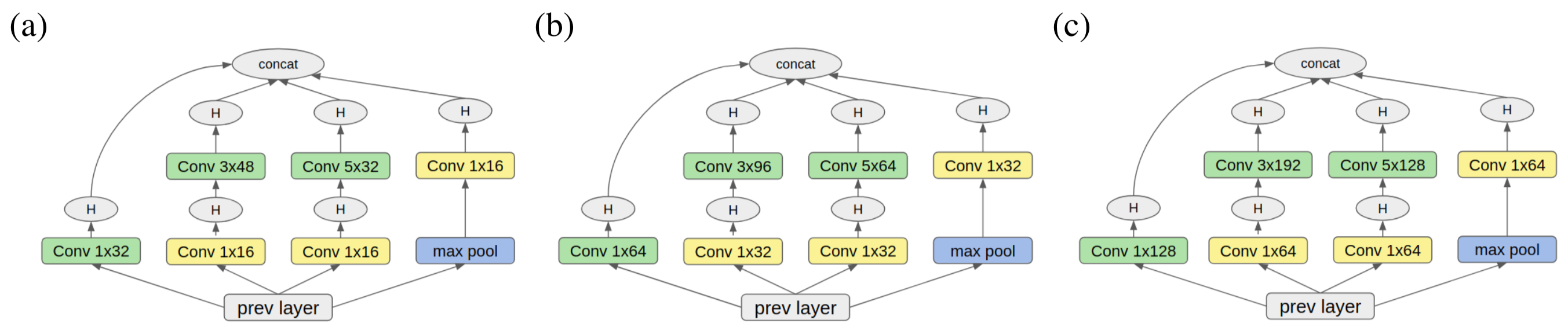}
            	\centering
            	\caption{(a) inception A. (b) inception B. (c) inception C. The inception block used in our method (refer to table 1). The input feature maps are passed through 4 different paths with convolution having different filter size and max pooling. The convolution with the yellow background is viewed as dimension reduction.}
            \end{figure}

        \subsubsection{Inception Block}
            Because of the limited number of manually annotated amino acid sequences, we choose the parameter-wise efficient network architecture.
            Our inception is based on the 1D convolution neural network, and the architecture of inception block is shown in Fig. 5. 

\subsection{Loss Function}
We use cross entropy and L2 weight regularization as our loss function. The loss function can be described as below
\begin{equation}
\mathbf{L_{e}}(\mbox{S}, \phi) = \frac{1}{N} \times \sum_{i=1}^{q}y_i \log \hat{y_{i}} + (1-y_{i})\log(1-\hat{y_{i}})
\end{equation}

\begin{equation}
\mathbf{L_{r}}(\phi) = \sum_{weight \in \phi}^{} weight^{2}
\end{equation}

\begin{equation}
\mathbf{L}(\mbox{S}, \phi) = \mathbf{L_{e}}(\mbox{S}, \phi) + \alpha \cdot \mathbf{L_{r}}(\phi)
\end{equation}
where $\hat{y_{i}} \in$ \{0, 1\} and $q$ are the ground truth label and the number of GO labels. As $\hat{y_{i}}$ = 1, it indicates that the sequence has a function $\hat{y_{i}}$. $y_{i}$ $\in$ $Y$ is the prediction result of our method. $\alpha$ is the scalar of L2 weight regularization loss.

\section{Experiment}
    We use the protein sequence data from UniProtKB/Swiss-Prot 3/9/2018 version and their corresponding GO annotations. The total number of the amino acid sequences is 530529. Besides, there are 44939 GO terms. The number of the three domains of GO (i.e., CC, MF and BP) are 4171, 11154, and 29614, respectively. Besides, to train word embedding, we collect 1500000 amino acid sequences form UniProtKB/TrEMBL whose functions aren't annotated. We will introduce the training process in Sec. 5.1, metrics in Sec. 5.2, visualization of learned weights in ELMo in Sec. 5.3 and inference speed in Sec. 5.4.

    \subsection{Training}
        \subsubsection{Word Embedding}
            We use ELMo word representation as our word embedding. Besides, we use Skip-Gram word embedding which is widely used in natural language processing as our baseline. \par
            We train the ELMo word representation with Adagrad optimizer \cite{duchi2011adaptive}. The dimension of the language model in ELMo is 256. After training with 10 epochs, we freeze the weights in the language model in ELMo.\par
            For Skip-Gram word embedding \citep{NIPS2013_5021}, we train Skip-Gram with Adam optimizer \cite{journals/corr/KingmaB14}. After training with 13 epochs, we freeze the weights in the Skip-Gram model and regard the matrix in Skip-Gram as embedding lookup table. For the hyperparameters of word embedding, please refer to supplement material.

        \subsubsection{Amino Acid Sequence Model}
            All the models are trained with Adam optimizer. On the UniProtKB/Swiss-Prot dataset, we train our model with batch size 64 and 12 epochs and apply grid search to find the best hyperparameters (see supplement material). For convenience consideration, we fix the length of each amino acid sequence to 608 tokens, i.e. 1824 amino acids since a token repersent 3 amino acids. If the sequence is shorter than 608 tokens, we pad the sequence until 608 tokens. On the other hand, if the sequence is longer than 608 words, we keep the first 608 tokens and remove the remaining tokens. The amino acid sequences below 1824 amino acids (608 tokens) account for 99.28\% of our dataset.

    \subsection{Metrics}
        We measure the performance with F1 scores. To calculate F1 scores, we need to measure precision and recall. 

        \begin{equation}
        F1   Score = \frac{2 \times Precision \times Recall}{(Precision + Recall)}
        \end{equation}

    \begin{table*}
    	\centering
    	\resizebox{\linewidth}{!}{
        \begin{tabular}{llrrrr}
            \hline
            \multicolumn{4}{r}{F1 Scores (mean/variance)} \\
            \cline{2-5}
            Model    & Total & BP & CC & MF \\
            \hline
            BLAST  &  - & $0.31^{*}$ & $0.37^{*}$ & $0.36^{*}$     \\ \hline
            DeepGO        & - & $0.36^{*}$ & $0.46^{*}$ & $0.63^{*}$      \\ \hline
            PANNZER2		& - & $0.699^{+}$ & $0.823^{+}$ & $0.708^{+}$	\\ \hline \hline
            Skip-Gram+inception & $0.767/2.4\times 10^{-6}$ & $0.752/4.1\times 10^{-6}$ & $0.780/5.2\times 10^{-6}$ & $0.764/1.1\times 10^{-6}$      \\ \hline
            \textbf{ELMo+inception} & $\mathbf{0.835}/2.1\times 10^{-6}$ &  $\mathbf{0.813}/3.1\times 10^{-6}$ &  $\mathbf{0.842}/4.9\times 10^{-6}$ & $\mathbf{0.854}/1.6\times 10^{-5}$ \\
            \hline
        \end{tabular}}
        
      	\caption{Performance comparison. * indicates that the performance is referred DeepGO. + indicates that the performance is referred from PANNZER2. Note that PANNZER2 is the previous state-of-the-art method. Since the performance of our method is measured by cross-validation on 5 equal folds, we show both the mean and variance of F1 score on the table.}
    \end{table*}

    In table 1, we compare the performance in different models. Obviously, inception with ELMo can reach the best performance. Note that the performances of our model are based on 5-fold cross validation and the performances of other methods are referred from their paper. We show the mean and variance of F1 scores in table 1. Besides, since the configurations of DeepGO are slightly different between difference methods (e.g. DeepGO uses a self-defined subset of GO), we also provide the performance which measure on exactly the same testing set and GO version in table 2. 
    
    \begin{table}
    	\centering
    
        \begin{tabular}{lrrr}
            \hline
            \multicolumn{4}{c}{F1 Scores} \\
            \cline{2-4}
            Model     & BP & CC & MF \\
            \hline
            BLAST   & $0.31^{*}$ & $0.37^{*}$ & $0.36^{*}$     \\ \hline
            DeepGO  & $0.36^{*}$ & $0.46^{*}$ & $0.63^{*}$      \\ \hline 
            \textbf{ELMo+inception}  & $\mathbf{0.8019}$ &  $\mathbf{0.7066}$ & $\mathbf{0.8832}$ \\ \hline
        \end{tabular}
        
      	\caption{Performance comparison based on exactly the same test set and GO version of DeepGO experiments. We can see that our method is better than other method. * indicates that the performance is referred DeepGO}
    \end{table}

        \subsection{Visualization of Learned Weights in ELMo}
            To further explore the effects of ELMo in our method, we train our model in a different way. We separate the label space into three sub-spaces according to the three domains of GO, i.e. BP, CC, and MF. Then, we visualize the softmax-normalized weights in the language model of ELMo (see Fig. 6). \par 
            
            Obviously, we can observe that all the three domains of GO use fewer features from the first layer of the language model in ELMo than those from other layers. Besides, MF relies more on the features from the second layer than that from the third layer. As we illustrated in the background, MF is defined as the functionalities of a gene product at the molecular level. This is usually formed by subsequences with specific amino acid arrangement patterns. BP is a recognized series of collaborated MF terms or chemical reactions. CC marks where a gene product locates.  \par
            
            By extracting the features from amino acid sequences, the low-level features that we can obtain are highly dependent on the amino acid sequence arrangement pattern. Since MF has the closest relationship to that, we can conclude that MF would have its features extracted first. BP is organized from the MF terms or molecular reactions, hence BP would have its features extracted after those of MF. Identifying the location requires the overall information about the protein function, hence features of CC tend to be extracted last. Due to these cause-effect relations between the three GO domains, we can explain the distribution of weights in the language model of ELMo. In conclusion, the later the features are extracted, the harder the task relies on a deeper layer, and vice versa.
            
            \begin{figure}[h]
            	\includegraphics[scale=0.23]{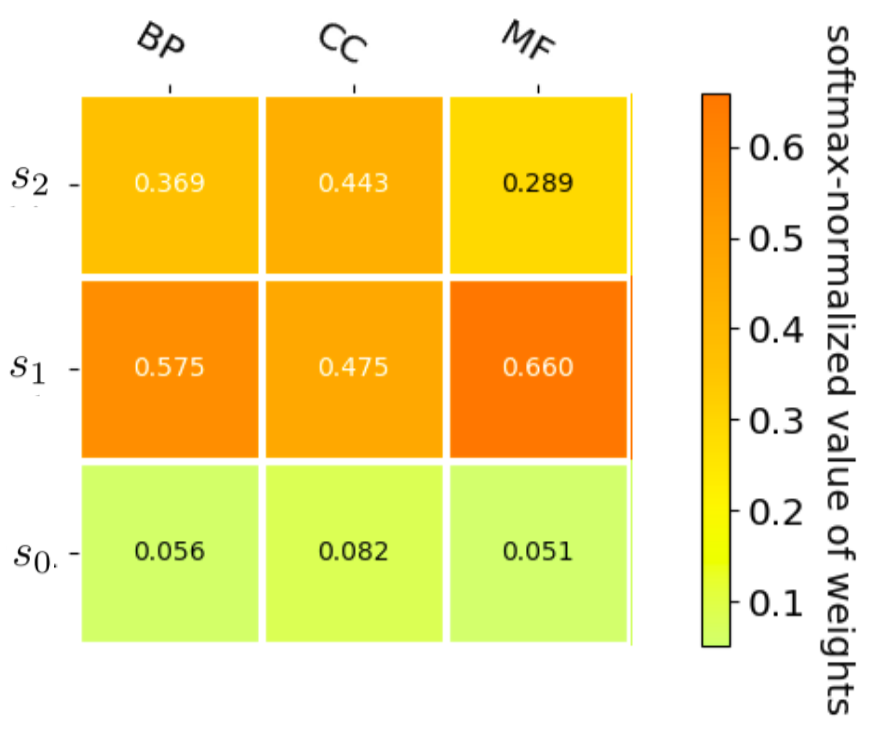}
            	\centering
            	\caption{The weights in the language model of ELMo for different domains of GO. $s_{0}$, $s_{1}$ and $s_{2}$ are the softmax-normalized weights in the first, the second and the third layer.}
            \end{figure}

    \subsection{Inference Speed}
        The median of the amino acid sequence length is 294. Therefore, we select the amino acid sequence that has 294 amino acids from UniProtKB/Swiss-Prot to evaluate the inference speed. Then, we attempt to measure the average time it takes to infer the functionalities of an amino acid sequence. The outcomes are shown in table 3.

        \begin{table}
        	\centering
        	\begin{tabular}{ | l | r | } 
        		\hline
                model & average inference time \\
                \hline
                \hline
        		BLAST&  3.65 s\\ 
                \hline
                DeepGO & 6.62 s\\
                \hline
                PANNZER2 & 2.067 s\\
                \hline
                ELMo+inception network & \textbf{0.140} s \\
                \hline
        	\end{tabular}
        	\caption{The average inference times (in seconds) of different models.  Our proposed method is significantly faster than other methods.}
        \end{table}

\section{Conclusion}
    We have introduced an approach to predict the functions and properties of the amino acid sequences. The deep learning model is a feasible solution to accelerate the process of confirming the amino acid sequence properties in biotechnology industry applications. Compared with the other methods, our method is more accurate and faster. We also show that word embedding in the amino acid sequences can efficiently extract the features in the early stage of the training process. 
    
\medskip
\small
\bibliographystyle{unsrt}
\bibliography{thesis}

\begin{thebibliography}{10}

\bibitem{doi:10.1093/nar/gkw1099}
TheUniProtConsortium.
\newblock Uniprot: the universal protein knowledgebase.
\newblock {\em Nucleic Acids Research}, 45(D1):D158--D169, 2017.

\bibitem{Ashburner2000}
Michael Ashburner, Catherine~A. Ball, Judith~A. Blake, David Botstein, Heather
  Butler, J.~Michael Cherry, Allan~P. Davis, Kara Dolinski, Selina~S. Dwight,
  Janan~T. Eppig, Midori~A. Harris, David~P. Hill, Laurie Issel-Tarver, Andrew
  Kasarskis, Suzanna Lewis, John~C. Matese, Joel~E. Richardson, Martin
  Ringwald, Gerald~M. Rubin, and Gavin Sherlock.
\newblock Gene ontology: tool for the unification of biology.
\newblock {\em Nature Genetics}, 25:25--29, May 2000.

\bibitem{doi:10.1093/nar/gkw1108}
TheGeneOntologyConsortium.
\newblock Expansion of the gene ontology knowledgebase and resources.
\newblock {\em Nucleic Acids Research}, 45(D1):D331--D338, 2017.

\bibitem{Peters:2018}
Matthew~E. Peters, Mark Neumann, Mohit Iyyer, Matt Gardner, Christopher Clark,
  Kenton Lee, and Luke Zettlemoyer.
\newblock Deep contextualized word representations.
\newblock In {\em Proc. of NAACL}, 2018.

\bibitem{szegedy2015going}
Christian Szegedy, Wei Liu, Yangqing Jia, Pierre Sermanet, Scott Reed, Dragomir
  Anguelov, Dumitru Erhan, Vincent Vanhoucke, and Andrew Rabinovich.
\newblock Going deeper with convolutions.
\newblock In {\em Proceedings of the IEEE conference on computer vision and
  pattern recognition}, pages 1--9, 2015.

\bibitem{BLAST}
Altschul SF1, Gish W, Miller W, Myers EW, and Lipman DJ.
\newblock Basic local alignment search tool.
\newblock {\em NICB}, 1990.

\bibitem{doi:10.1093/bioinformatics/btu851}
Patrik Koskinen, Petri Törönen, Jussi Nokso-Koivisto, and Liisa Holm.
\newblock Pannzer: high-throughput functional annotation of uncharacterized
  proteins in an error-prone environment.
\newblock {\em Bioinformatics}, 31(10):1544--1552, 2015.

\bibitem{doi:10.1093/nar/gky350}
Petri Törönen, Alan Medlar, and Liisa Holm.
\newblock Pannzer2: a rapid functional annotation web server.
\newblock {\em Nucleic Acids Research}, 46(W1):W84--W88, 2018.

\bibitem{kulmanov2017deepgo}
Maxat Kulmanov, Mohammed~Asif Khan, and Robert Hoehndorf.
\newblock Deepgo: predicting protein functions from sequence and interactions
  using a deep ontology-aware classifier.
\newblock {\em Bioinformatics}, 34(4):660--668, 2017.

\bibitem{NIPS2013_5021}
Tomas Mikolov, Ilya Sutskever, Kai Chen, Greg~S Corrado, and Jeff Dean.
\newblock Distributed representations of words and phrases and their
  compositionality.
\newblock In C.~J.~C. Burges, L.~Bottou, M.~Welling, Z.~Ghahramani, and K.~Q.
  Weinberger, editors, {\em Advances in Neural Information Processing Systems
  26}, pages 3111--3119. Curran Associates, Inc., 2013.

\bibitem{DBLP:journals/corr/KirosZSZTUF15}
Ryan Kiros, Yukun Zhu, Ruslan Salakhutdinov, Richard~S. Zemel, Antonio
  Torralba, Raquel Urtasun, and Sanja Fidler.
\newblock Skip-thought vectors.
\newblock {\em CoRR}, abs/1506.06726, 2015.

\bibitem{logeswaran2018an}
Lajanugen Logeswaran and Honglak Lee.
\newblock An efficient framework for learning sentence representations.
\newblock In {\em International Conference on Learning Representations}, 2018.

\bibitem{DBLP:journals/corr/SimonyanZ14a}
Karen Simonyan and Andrew Zisserman.
\newblock Very deep convolutional networks for large-scale image recognition.
\newblock {\em CoRR}, abs/1409.1556, 2014.

\bibitem{DBLP:journals/corr/HeZRS15}
Kaiming He, Xiangyu Zhang, Shaoqing Ren, and Jian Sun.
\newblock Deep residual learning for image recognition.
\newblock {\em CoRR}, abs/1512.03385, 2015.

\bibitem{DBLP:journals/corr/HuangLW16a}
Gao Huang, Zhuang Liu, and Kilian~Q. Weinberger.
\newblock Densely connected convolutional networks.
\newblock {\em CoRR}, abs/1608.06993, 2016.

\bibitem{DBLP:journals/corr/JozefowiczVSSW16}
Rafal J{\'{o}}zefowicz, Oriol Vinyals, Mike Schuster, Noam Shazeer, and Yonghui
  Wu.
\newblock Exploring the limits of language modeling.
\newblock {\em CoRR}, abs/1602.02410, 2016.

\bibitem{DBLP:journals/corr/KimJSR15}
Yoon Kim, Yacine Jernite, David Sontag, and Alexander~M. Rush.
\newblock Character-aware neural language models.
\newblock {\em CoRR}, abs/1508.06615, 2015.

\bibitem{Ioffe2015BatchNA}
Sergey Ioffe and Christian Szegedy.
\newblock Batch normalization: Accelerating deep network training by reducing
  internal covariate shift.
\newblock In {\em ICML}, 2015.

\bibitem{DBLP:journals/corr/ClevertUH15}
Djork{-}Arn{\'{e}} Clevert, Thomas Unterthiner, and Sepp Hochreiter.
\newblock Fast and accurate deep network learning by exponential linear units
  (elus).
\newblock {\em CoRR}, abs/1511.07289, 2015.

\bibitem{duchi2011adaptive}
John Duchi, Elad Hazan, and Yoram Singer.
\newblock Adaptive subgradient methods for online learning and stochastic
  optimization.
\newblock {\em Journal of Machine Learning Research}, 12(Jul):2121--2159, 2011.

\bibitem{journals/corr/KingmaB14}
Diederik~P. Kingma and Jimmy Ba.
\newblock Adam: A method for stochastic optimization.
\newblock {\em CoRR}, abs/1412.6980, 2014.

\end{thebibliography}
\end{document}